\documentclass[mathleft]{an}
\usepackage{graphicx}
\usepackage{times}
\begin{document}

\Pagespan{1}{}
\Yearpublication{2010}%
\Yearsubmission{2010}%
\Month{Month}%
\Volume{Volume}%
\Issue{Issue}%

\title{On the Nature of the Purported Common Proper Motion Companions to the Exoplanet Host Star 51 Peg}

\author{Eric E. Mamajek}
\titlerunning{51 Peg}
\authorrunning{E. E. Mamajek}
\institute{Department of Physics \& Astronomy, University of Rochester, Rochester, NY 14627-0171 USA}

\received{29 March 2010}
\accepted{26 April 2010}
\publonline{}


\abstract{Greaves (2006) proposed that three red, high proper motion 
stars within 10$^{\circ}$ of 51 Peg (NLTT 54007, 54064, \& 55547) are
co-moving companions to this famous exoplanet host star. While the
stars clearly have proper motions similar to 51 Peg, the inferred
kinematic parallaxes for these stars produce extremely inconsistent
color-magnitude positions 2 to 4 magnitudes below the main
sequence. All three stars are likely to be background stars unrelated
to 51 Peg.}

\maketitle

\section{Introduction}

In October 1995, Mayor \& Queloz (1995) reported the discovery of a
Jovian-mass companion orbiting the solar-type star 51 Peg in a 4.2 day
orbit. The companion 51 Peg b is the prototype of the 'hot Jupiter'
class, and the 51 Peg system has been the source of intense study over
the past decade and a half. Thus far, surveys have failed to identify
any reliable stellar companions, either within a few arcseconds
(Luhman \& Jayawardhana 2002), and out to $\sim$5' (Raghavan et
al. 2006). The existence of low-mass stellar companions to exoplanet
host stars are of dynamical interest in the quest to understand the
diversity of planetary systems (e.g. Desidera \& Barbieri 2007).

Greaves (2006) reported that three faint stars might be co-moving with
51 Peg: NLTT 55547 (1$^{\circ}$.1 away from 51 Peg), NLTT 54064
(7$^{\circ}$.1) and NLTT 54007 (8$^{\circ}$.4). It is extremely
unlikely a priori that any of these stars would be {\it bound}
companions to 51 Peg as the maximum observed separation for $\sim$1
M$_{\sun}$ stars is in the range $\sim$2500-10000 AU
($\sim$0.012-0.048 pc; Abt 1988, Close et al. 2003), and the projected
separation of the nearest of these (NLTT 55547) is 0.28 pc
($\sim$58000 AU) if codistant with 51 Peg. The association of these
stars with 51 Peg predicates not only on the proper motions of these
stars, but also their color- magnitude data (and of course parallax
and radial velocity). In this contribution I conclude that for all
three stars the color-magnitude data combined with the kinematic
distances (predicted by combining their proper motions and the space
velocity of 51 Peg) are sufficient to rule out companionship to 51
Peg.

\section{Analysis}

\subsection{51 Peg}

Astrometric data for 51 Peg are presented in Table 1, with the
position, proper motion and parallax coming from van Leeuwen
(2007). Longer baseline proper motions by Hog et al. (2000) and Roeser
et al. (2008) agree within 2$\sigma$ along each component. The
systemic radial velocity for 51 Peg comes from Nidever et al. (2002),
and no long-term radial velocity trend for 51 Peg was reported. Since
the Sun and 51 Peg are of nearly identical spectral type, and the zero
point of the radial velocity is calibrated using the Sun (via
measurements of radial velocities of minor planets), the systemic
radial velocity of 51 Peg should accurately correlate to the true
radial velocity within $\sim$0.1 km\,s$^{-1}$ (which is the
uncertainty I adopt; see discussion by Nidever et al. 2002). Hence,
the galactic velocity estimated and quoted in Table 1 should
accurately represent the center-of-mass velocity of the 51 Peg system
to within $\pm$0.1 km\,s$^{-1}$ with respect to the solar system's
barycenter. The velocity vector in Table 1 is on the Galactic
Cartesian system where $U$ is directed towards the Galactic center,
$V$ is in the direction of Galactic rotation, and $W$ is towards the
North Galactic pole. This velocity vector is used to calculate the
vertex of 51 Peg - i.e. the point on the celestial sphere where the
proper motion vector of any star with identical velocity will point
(essentially the ``convergent point'' for a single star rather than an
entire cluster; e.g. Atanasijevic 1971). The vertex position in Table
1 is accurate to $\pm$0$^{\circ}$.13.

51 Peg is an old Population I G-dwarf with an isochronal age of
6.8\,$\pm$\,1.5 Gyr (Takeda et al. 2007) and chromospheric activity
age similarly in the range of 6.1-8.1 Gyr (Mamajek \& Hillenbrand
2008). 51 Peg is also metal-rich ([Fe/H] = 0.20; Valenti \& Fischer
2005), so not only would any physical companion have a velocity
similar to that of 51 Peg, but it would be expected to be of the same
age and metal-rich as well.

\begin{table}
\caption{Properties of 51 Peg}
\label{table:properties}
\begin{tabular}{lll}
\hline
\multicolumn{1}{c}{Property} &
\multicolumn{1}{c}{Value} &
\multicolumn{1}{c}{Ref.}\\
\hline
$\alpha$(ICRS, J1991.25)        & 344$^{\circ}$.36604474 & 1\\
$\delta$(ICRS, J1991.25)        & +20$^{\circ}$.76868422 & 1\\
Parallax                        & 64.07\,$\pm$\,0.38 mas & 1\\
Distance                        & 15.61\,$\pm$\,0.09 pc  & 2\\
$\mu_{\alpha}$                  & 207.25\,$\pm$\,0.31 mas\,yr$^{-1}$ & 1\\
$\mu_{\delta}$                  &  60.34\,$\pm$\,0.30 mas\,yr$^{-1}$ & 1\\
Radial Velocity                 & --33.225 km\,s$^{-1}$  & 3\\
Velocity ($U$)                  & --15.40\,$\pm$\,0.09 km\,s$^{-1}$ & 2\\
Velocity ($V$)                  & --29.66\,$\pm$\,0.07 km\,s$^{-1}$ & 2\\
Velocity ($W$)                  &  +15.56\,$\pm$\,0.08 km\,s$^{-1}$ & 2\\
Speed ($S$)                     & 36.86\,$\pm$\,0.08 km\,s$^{-1}$ & 2\\
Vertex ($\alpha$)               & 139$^{\circ}$.21\,$\pm$\,0$^{\circ}$.13 & 2\\ 
Vertex ($\delta$)               & --11$^{\circ}$.91\,$\pm$\,0$^{\circ}$.12 & 2\\
\hline \\
\end{tabular}
\begin{flushleft}
References: (1) van Leeuwen (2007), (2) calculated by author using other data in
table, (3) Nidever et al. (2002).
\end{flushleft}
\end{table}

\subsection{NLTT 55547}

NLTT 55547 (= LP 401-32 = 2MASS J23003379+2135223) is situated
1$^{\circ}$.03 from 51 Peg. Its position, proper motion, and relevant
photometry are listed in Table
\ref{tab:NLTT_55547}. There is no published spectral type
for the star, with Luyten (1979) simply suggesting that the star was
``k-m'' type based on color. There are multiple estimates of a V
magnitude based on photographic plate scans: 15.93 (Lepine \& Shara
2005), 15.61 (Salim et al. 2003), 15.74\,$\pm$\,0.34 (Lasker et
al. 2008), and I estimate 15.48 based on combining the USNO A2.0 B and
R magnitudes. I adopt a median V magnitude of 15.68 based on these 4
estimates. The (V-Ks) color (4.58) is consistent with the mean color
of M3V stars in the CNS3 catalog (Gliese \& Jahreiss 1991) using
photometry compiled by Neill
Reid\footnote{http://www.stsci.edu/$\sim$inr/cmd.html}. The (H-K$_s$) color
is consistent with being in the range of M1V-M3.5V (1$\sigma$ range),
using median colors of M dwarfs compiled by the author by
cross-referencing 2MASS photometry (Cutri et al. 2003) with M dwarfs
classified on the Kirkpatrick-Henry system of standards in the Dwarf
Archives database (Gelino, Kirkpatrick, \& Burgasser
2009)\footnote{http://spider.ipac.caltech.edu/staff/davy/ARCHIVE/index.shtml}. I
conclude that a photometric estimate of the star's spectral type is
$\sim$M3V.

I estimate the proper motion for this star using the same positions
used by Lepine \& Shara (2005), and recover the proper motion quoted
by Lepine \& Shara to within 0.5 mas\, yr$^{-1}$. I adopt their proper
motion value with an estimated uncertainty of 4 mas\, yr$^{-1}$. I
follow the techniques used in Mamajek et al. (2002) and Mamajek (2005)
to rotate the equatorial proper motion into components pointed towards
the 51 Peg vertex ($\mu_{\upsilon}$) and perpendicular
($\mu_{\tau}$). One can also calculate a predicted cluster parallax
distance (i.e. a ``kinematic distance''), and combine that predicted
distance with the $\mu_{\tau}$ to estimate a peculiar velocity
(i.e. an estimate of the velocity of the star perpendicular to the
component of its velocity directed towards the 51 Peg vertex). I
rotate the equatorial proper motion for NLTT 55547 into the proper
motion components towards the 51 Peg vertex ($\mu_{\upsilon}$ =
210\,$\pm$\,4 mas\, yr$^{-1}$) and perpendicular ($\mu_{\tau}$ =
5\,$\pm$\,4 mas\, yr$^{-1}$). If NLTT 55547 has the same 3D velocity
as 51 Peg, its proper motion toward the 51 Peg vertex (situated
153$^{\circ}$.4 away) suggests a distance of 16.6\, $\pm$\, 0.3 pc
(parallax $\varpi$ = 60.2\,$\pm$\,1.2 mas), and peculiar velocity
$v_{pec}$ = 0.4\, $\pm$\, 0.3 km\,s$^{-1}$. This solution predicts
that NLTT 55547 would have a radial velocity of -33.0\, $\pm$\, 0.1
km\,s$^{-1}$. If the kinematic distance is correct than the true
separation from 51 Peg is 1.0 pc, considerably more than the projected
separation of 0.28 pc if NLTT 55547 were codistant with 51 Peg. An
object at the celestial position of NLTT 55547, but with identical
distance and velocity vector as 51 Peg, would have a proper motion of
$\mu_{\alpha}$, $\mu_{\delta}$ = 213.1, +65.7 mas\, yr$^{-1}$
(i.e. nearly identical $\mu_{\delta}$, but $\mu_{\alpha}$ deviant by
3.5$\sigma$). An object at the celestial position of NLTT 55547, with
the velocity vector of 51 Peg, but distance of 16.6 pc, would have a
proper motion of $\mu_{\alpha}$, $\mu_{\delta}$ = 200.4, +61.8 mas\,
yr$^{-1}$ - i.e.  within 4 mas\, yr$^{-1}$ and 1$\sigma$ of the
observed proper motion components.  So to force the projected motion
of NLTT 55547 to match the predictions for a star comoving with 51
Peg, NLTT 55547 would have to be slightly further away (16.6 pc) than
51 Peg (15.6 pc).

\begin{figure}
\includegraphics[width=80mm,height=80mm]{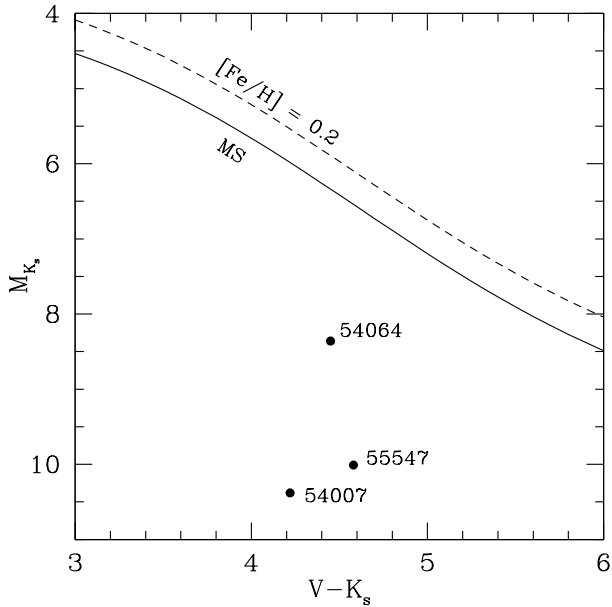}
\caption{V-K$_s$ color vs. absolute magnitude M$_{Ks}$ diagram. The main sequence
of Johnson \& Apps (2009) is shown as a {\it solid line}, which they
claim corresponds to the mean metallicity of field M dwarfs ([Fe/H] =
-0.05). Their sequence for [Fe/H] = +0.2 (the metallicity of 51 Peg)
is shown as a {\it dashed line}. The color-magnitude positions of the
NLTT stars at their kinematic distances (assuming they share the space
velocity of 51 Peg) are plotted as {\it filled circles}.}
\label{fig:vk_mk}
\end{figure}

Does NLTT 55547's photometric data seem consistent with the distance
predicted by its potential co-motion with 51 Peg? At the kinematic
distance of 16.6\, $\pm$\, 0.3 pc, NLTT 55547 would have an absolute
magnitude of M$_{Ks}$ = 10.01\, $\pm$\, 0.05. I determined the range
of plausible absolute K$_s$ magnitudes for M dwarfs of (V-K$_s$)
$\simeq$ 4.57 using Figure 2 of Johnson \& Apps (2009). Johnson \&
Apps (2009) also provide an absolute magnitude offset as a function of
[Fe/H]. The predicted $M_{K_s}$ for the main sequence for (V-K$_s$) =
4.57 is $M_{K_s}$ = 6.54, and correcting for the slightly higher
metallicity of 51 Peg ([Fe/H] = 0.20), one predicts an absolute
magnitude of 6.10 (see Figure
\ref{fig:vk_mk}). Hence, if NLTT 55547 lies at the kinematic distance
of 16.6 pc (consistent with comoving with 51 Peg), then {\it the
predicted absolute K$_s$ magnitude is 4 magnitudes too faint for its
(V-K$_s$) color}. The conclusion is obviously the same if one simply
adopts the distance of 51 Peg for NLTT 55547. While Greaves (2006)
states that the derived ``absolute magnitude value is not
inappropriate for a cooler spectral class M dwarf'', the colors are
clearly consistent with an early ($\sim$M3) dwarf. Typical stars with
M$_{K_s}$ $\simeq$ 10 have (V-K$_s$) colors of $\sim$8 to $\sim$10,
hence no reasonable amount of photometric error could explain the
offset.  If NLTT 55547 is a typical mid-M dwarf, its (V-K$_s$) and
K$_s$ data are consistent with a distance modulus of $\sim$4.6 mag, a
photometric distance of $\sim$80 pc, and inferred tangential velocity
of $\sim$79 km\,s$^{-1}$.

I conclude that despite the remarkable match of proper motion between
NLTT 55547 and 51 Peg, the color- magnitude data are completely
inconsistent with the two stars being physically related.

\begin{table}
\caption{Properties of NLTT 55547}
\label{tab:NLTT_55547}
\begin{tabular}{lll}
\hline
Property & Value & Ref.\\
\hline
$\alpha$(ICRS, J2000) & 345$^{\circ}$.140900 & 1\\
$\delta$(ICRS, J2000) & +21$^{\circ}$.589565 & 1\\
$\mu_{\alpha}$        & 199\,$\pm$\,4 mas\, yr$^{-1}$ & 1\\
$\mu_{\delta}$        & 65\,$\pm$\,4 mas\, yr$^{-1}$ & 1\\
V mag                 & 15.68                & 1\\
J mag                 & 11.957\,$\pm$\,0.023 & 2\\
H mag                 & 11.362\,$\pm$\,0.022 & 2\\
Ks mag                & 11.114\,$\pm$\,0.018 & 2\\
(J-H)                 & 0.595\,$\pm$\,0.032  & 2\\
(H-Ks)                & 0.248\,$\pm$\,0.028  & 2\\
(V-Ks)                & 4.57                 & 1,2\\
(J-Ks)                & 0.843\,$\pm$\,0.029  & 2\\
$\mu_{\upsilon}$      & 210\,$\pm$\,4 mas\, yr$^{-1}$ & 3\\
$\mu_{\tau}$          &   5\,$\pm$\,4 mas\, yr$^{-1}$ & 3\\
$v_{pec}$             & 0.4\,$\pm$\,0.3 km\,s$^{-1}$ & 3\\
Kinematic Distance    & 16.6\,$\pm$\,0.3 pc & 3\\
Predicted RV          & -33.0\,$\pm$\,0.1 km\,s$^{-1}$ & 3\\
Predicted M$_{Ks}$    & 10.01\,$\pm$\,0.05 & 3\\
Predicted M$_V$       & 14.6               & 3\\
Photometric Distance  & $\sim$80 pc        & 3\\
\hline 
\end{tabular}
\begin{flushleft} References: (1) Lepine \& Shara (2005), (2) Cutri et al. (2003), 
(3) this work. Kinematic distance and predicted values assume that the star shares the 3D velocity of 51
Peg.
\end{flushleft}
\end{table}

\subsection{NLTT 54064}

NLTT 54064 (= LP 460-37 = 2MASS J22311528+1709444) is situated
7$^{\circ}$.17 from 51 Peg. Its position, proper motion, and relevant
photometry is listed in Table \ref{tab:NLTT_54064}. There is no
published spectral type for NLTT 54064, however Luyten (1979) surmised
that it was ``m'' class based on its colors. Several authors list V
magnitudes: Rapaport et al. (2001) report 13.097\, $\pm$\, 0.058,
Ducourant et al. (2006) report 13.262\, $\pm$\, 0.077, Droege et
al. (2007) report 12.415\, $\pm$\, 0.144, and Lepine \& Shara (2005)
list 13.15.  I adopt a median V magnitude of 13.12. The star's V-K$_s$
and H-K$_s$ colors are consistent with an $\sim$M2 dwarf. Lepine \&
Shara (2005) estimate a photometric distance of 32.5\, $\pm$\, 9.8
pc. The proper motion estimates for NLTT 54064 across several catalogs
agree within their errors (Salim et al. 2003, Lepine
\& Shara 2005, Ducourant et al. 2006, Roeser et al. 2008). I adopt the
UCAC2 proper motion as the best available (Zacharias et al. 2004).

The proper motion of NLTT 54064 directed towards the vertex of 51 Peg
is $\mu_{\upsilon}$ = 216.0\, $\pm$\, 4.2 mas\, yr$^{-1}$, and the
perpendicular component is $\mu_{\tau}$ = 16.2\, $\pm$\, 4.2 mas\,
yr$^{-1}$. The predicted kinematic distance (assuming NLTT 54064
shares the same velocity as 51 Peg) is 11.6\, $\pm$\, 0.2 pc (parallax
$\varpi$ = 86.5\, $\pm$\, 1.8 mas), and the predicted radial velocity
is -34.9 km\,s$^{-1}$. Given the predicted kinematic distance and
perpendicular proper motion, the peculiar velocity is 0.9\, $\pm$\,
0.2 km\,s$^{-1}$ (which corresponds to the {\it minimum} velocity
difference between NLTT 54064 and 51 Peg).

At the kinematic distance 11.6 pc, NLTT 54064 would have an absolute
magnitude of M$_{Ks}$ = 8.36\, $\pm$\, 0.05, some 2.0 magnitudes below
the main sequence, and 2.5 mag below the [Fe/H] = 0.2 V-K$_s$
vs. M$_{Ks}$ sequence of Johnson \& Apps (2009) (Figure
\ref{fig:vk_mk}).  If NLTT 54064 is a field M dwarf unassociated with 51 Peg,
I estimate a photometric distance of $\sim$34 pc for NLTT 54064,
nearly identical to that estimated by Lepine \& Shara (2005) (32.5
pc). At this distance the star would have a tangential velocity of
$\sim$35 km\,s$^{-1}$. Similar to the first star, I find that while
NLTT 54064 shows {\it projected} motion similar to 51 Peg (within
$\sim$1 km\,s$^{-1}$ peculiar velocity), its color-magnitude position
at the predicted kinematic distance ($\sim$12 pc) is completely
inconsistent with being a [Fe/H] = 0.2 M-dwarf.

\begin{table}
\caption{Properties of NLTT 54064}
\label{tab:NLTT_54064}
\begin{tabular}{lll}
\hline
Property & Value & Ref.\\
\hline
$\alpha$(ICRS, J2000) & 337$^{\circ}$.813782 & 1\\
$\delta$(ICRS, J2000) & +17$^{\circ}$.16238  & 1\\
$\mu_{\alpha}$        & 206.0\,$\pm$\,4.2 mas\, yr$^{-1}$ & 2\\
$\mu_{\delta}$        & 67.1\,$\pm$\,4.2 mas\, yr$^{-1}$ & 2\\
V mag                 & 13.12 & 1\\
J mag                 & 9.518\,$\pm$\,0.022 & 4\\
H mag                 & 8.903\,$\pm$\,0.021 & 4\\
Ks mag                & 8.671\,$\pm$\,0.018 & 4\\
(J-H)                 & 0.615\,$\pm$\,0.031 & 4\\
(H-Ks)                & 0.232\,$\pm$\,0.028 & 4\\
(V-Ks)                & 4.45 & 3\\
(J-Ks)                & 0.847\,$\pm$\,0.029 & 3\\
$\mu_{\upsilon}$      & 216.0\,$\pm$\,4.2 mas\, yr$^{-1}$ & 3\\
$\mu_{\tau}$          & 16.2\,$\pm$\,4.2 mas\, yr$^{-1}$ & 3\\
$v_{pec}$             & 0.9\,$\pm$\,0.2 km\,s$^{-1}$ & 3\\
Kinematic Distance    & 11.6\,$\pm$\,0.2 pc & 3\\
Predicted RV          & -34.9 km\,s$^{-1}$ & 3\\
Predicted M$_{Ks}$    & 8.36\,$\pm$\,0.05 mag & 3\\
Predicted M$_V$       & 12.81 & 3\\
Photometric Distance  & $\sim$34 pc & 3\\
\hline 
\end{tabular}
\begin{flushleft}
References: (1) Lepine \& Shara (2005), (2) Zacharias et al. (2004), (3) calculated
by author from available data, (4) Cutri et al. (2003).
Kinematic distance and predicted values assume that the star shares the 3D velocity of 51 Peg. 
\end{flushleft}
\end{table}

\subsubsection{NLTT 54007}

NLTT 54007 (= LP 520-24 = 2MASS J22300000+1523483) is situated
8$^{\circ}$.45 from 51 Peg. Its position, proper motion, and relevant
photometry is listed in Table \ref{tab:NLTT_54007}. Luyten (1979)
considered the star 'k' type based on photographic colors, however no
spectroscopic type for the star has been reported in the literature. V
magnitudes of 14.884 (Rapaport et al. 2001), 14.926\, $\pm$\, 0.184
(Ducourant et al. 2006), 14.67\,$\pm$\,0.40 (Lasker et al. 2008),
15.04 (Salim et al. 2003) have been reported. I adopt a median V
magnitude of 14.91. The V-K$_s$ and H-K$s$ colors in Table 4 are both
consistent with an M2 dwarf. I adopt the proper motion for NLTT 54007
from Zacharias et al. (2004), which is similar to other motions with
larger uncertainties quoted by Lepine \& Shara (2005), Roeser et
al. (2008), \& Salim et al. (2003).
 
The proper motion of NLTT 54007 towards the vertex of 51 Peg is
$\mu_{\upsilon}$ = 209.3\, $\pm$\, 4.2 mas\, yr$^{-1}$, and the
perpendicular component is $\mu_{\tau}$ = 32.7\, $\pm$\, 4.2
mas\, yr$^{-1}$. If NLTT 54007 shares the space velocity of 51 Peg, its
proper motion is consistent with a kinematic distance of 11.5 $\pm$
0.2 pc (parallax 86.6 $\pm$ 1.9 mas), and one would predict a radial
velocity of -35.0 km\,s$^{-1}$. At the kinematic distance, the
perpendicular motion $\mu_{\tau}$ is consistent with a peculiar
velocity of 1.8 $\pm$ 0.2 km\,s$^{-1}$.

At the kinematic distance of 11.5 pc, NLTT 54007 would have an
absolute magnitude of M$_{Ks}$ = 10.38 $\pm$ 0.05. For its V-K$_s$
color, this is 4.4 magnitudes below the main sequence, and 4.81 mag
below the [Fe/H] = 0.2 V-K$_s$ vs. M$_{Ks}$ sequence of Johnson \&
Apps (2009) (see Figure \ref{fig:vk_mk}). If NLTT 54064 were a typical
field M dwarf, its colors suggest a photometric distance of $\sim$87
pc for NLTT 54064, and a probable tangential velocity of $\sim$87
km\,s$^{-1}$ at this distance. As with the other two stars, the
kinematic and color-magnitude data for NLTT 54007 are not consistent
with it being physically related to 51 Peg.

\begin{table}
\caption{Properties of NLTT 54007}
\label{tab:NLTT_54007}
\begin{tabular}{lll}
\hline
Property & Value & Ref.\\
\hline
$\alpha$(ICRS, J1991.25)  & 337$^{\circ}$.5001106 & 1\\
$\delta$(ICRS, J1991.25)  & +15$^{\circ}$.3967478 & 1\\
$\mu_{\alpha}$            & 201.8\,$\pm$\,4.2 mas\, yr$^{-1}$ & 1\\
$\mu_{\delta}$            & 64.4\,$\pm$\,4.2 mas\, yr$^{-1}$ & 1\\
V mag     & 14.91 & 2 \\
J mag     & 11.522\,$\pm$\,0.022 & 3\\
H mag     & 10.931\,$\pm$\,0.022 & 3\\
K$_s$ mag & 10.690\,$\pm$\,0.020 & 3\\
(J-H)                 & 0.591\,$\pm$\,0.031  & 3\\
(H-Ks)                & 0.241\,$\pm$\,0.030  & 2\\
(V-Ks)                & 4.22                 & 2,3\\
(J-Ks)                & 0.832\,$\pm$\,0.030  & 3\\
$\mu_{\upsilon}$      & 209.3\,$\pm$\,4.2 mas\, yr$^{-1}$ & 2\\
$\mu_{\tau}$          &  32.7\,$\pm$\,4.2 mas\, yr$^{-1}$ & 2\\
$v_{pec}$             &  1.8\,$\pm$\,0.2 km\,s$^{-1}$ & 2\\
Kinematic Distance    & 11.5\,$\pm$\,0.2 pc & 2\\
Predicted RV          & -35.0\,$\pm$\,0.1 km\,s$^{-1}$ & 2\\
Predicted M$_{Ks}$    & 10.38\,$\pm$\,0.05 & 2\\
Predicted M$_V$       & 14.60              & 2\\
Photometric Distance  & $\sim$87 pc & 2\\
\hline \\
\end{tabular}
\begin{flushleft} References: (1) Zacharias et al. (2004), (2) calculated by
author from available data,
(3) Cutri et al. (2003). Kinematic distance and predicted values assume that the star shares
the 3D velocity of 51 Peg.
\end{flushleft}
\end{table}

\section{Discussion}

There is one last possibility to consider for salvaging the hypothesis
that these stars could be co-moving with 51 Peg: the possibility that
these stars are white dwarfs. However none of the synthetic atmosphere
models for degenerate stars by Bergeron et al. (1995) produce objects
as red as these three stars (V-K$_s$ $\simeq$ 4.2-4.6), and indeed the
observed V-K$_s$ colors of both DA and non-DA white dwarfs are
generally less than V-K$_s$ $<$ 2.2 (Bergeron et al. 1997). Hence
there is no reason to believe a priori that the three stars could be
cool white dwarf companions to 51 Peg either.

I have demonstrated that while the proper motions of the three M
dwarfs are similar to that of 51 Peg, their inferred kinematic and
photometric distance estimates are very discordant, and hence none of
the stars are likely to be comoving stellar 'siblings' with 51 Peg.
Despite the lack of trigonometric parallax measurements for these
faint Luyten proper motion stars (NLTT 55547, 54064, and 54007), it
appears that the available color-magnitude and astrometric data are
probably sufficient to convincingly rule out Greave's (2006) claim of
physical association between these stars and 51 Peg. I estimate
photometric distances of $\sim$80, $\sim$34, and $\sim$87 pc for NLTT
55547, 54064, and 54007, respectively. The exercise demonstrates the
dangers of relying too heavily on proper motions alone on interpreting
the nature of widely separated stars of similar projected motion.

\acknowledgements

The author thanks Kevin Luhman for bringing the Greaves (2006) article
to his attention, and thanks he and Eric Bubar for commenting on the
manuscript. This publication makes use of data products from the Two
Micron All Sky Survey, which is a joint project of the University of
Massachusetts and the Infrared Processing and Analysis
Center/California Institute of Technology, funded by the National
Aeronautics and Space Administration and the National Science
Foundation. The Guide Star Catalog II is a joint project of the STScI
and the OATo. Space Telescope Science Institute is operated by the
Association of Universities for Research in Astronomy, for the
National Aeronautics and Space Administration under contract
NAS5-26555. The Osservatorio Astronomico di Torino is operated by the
Italian National Institute for Astrophysics (INAF). Additional support
was provided by the European Southern Observatory, Space Telescope
European Coordinating Facility, the International GEMINI project, and
the European Space Agency Astrophysics Division.

\end{document}